\documentclass[3p,times,twocolumn]{elsarticle}
 \biboptions{comma,sort&compress}
 
\usepackage{graphicx}
\usepackage{here}
\usepackage{ecrc}

\volume{00}

\firstpage{1}

\journalname{Nuclear and Particle Physics Proceedings}

\runauth{}

\jid{nppp}

\jnltitlelogo{Nuclear and Particle Physics Proceedings}

\usepackage{amssymb}

\usepackage[figuresright]{rotating}

\begin{document}

\begin{frontmatter}

\title{Short-distance HLbL contributions to the muon g-2 $^*$}
\cortext[cor0]{Talk given at 23rd International Conference in Quantum Chromodynamics (QCD 20),  27 October - 30 October 2020, Montpellier - FR}


\author[label1]{Johan Bijnens}
\ead{bijnens@thep.lu.se}
 
\author[label2]{Nils Hermansson-Truedsson\fnref{fn1}}
\ead{nils@itp.unibe.ch}

\author[label2]{Laetitia Laub}
\ead{laub@itp.unibe.ch}

\author[label1,label3]{Antonio Rodr\'{i}guez-S\'{a}nchez}
\ead{antonio.rodriguez\_sanchez@thep.lu.se, arodriguez@ijclab.in2p3.fr}

\address[label1]{Department of Astronomy and Theoretical Physics, Lund University, S\"{o}lvegatan 14A, SE 223 62 Lund, Sweden}

\address[label2]{Albert Einstein Center for Fundamental Physics, Institute for Theoretical Physics, Universit\"{a}t Bern, Sidlerstrasse 5, CH-3012 Bern, Switzerland}

\address[label3]{Present Address: Universit\'{e} Paris-Saclay, CNRS/IN2P3, IJCLab, 91405 Orsay, France}

\fntext[fn1]{Speaker, Corresponding author.}

\pagestyle{myheadings}
\markright{ }

\begin{abstract}
The current $3.7\sigma$ discrepancy between the Standard Model prediction and the experimental value of the muon anomalous magnetic moment could be a hint for the existence of new physics. The hadronic light-by-light contribution is one of the pieces requiring improved precision on the theory side, and an important step is to derive short-distance constraints for this quantity containing four electromagnetic currents. Here, we derive such short-distance constraints for three large photon loop virtualities and the external fourth photon in the static limit. The static photon is considered as a background field and we construct a systematic operator product expansion in the presence of this field. We show that the massless quark loop, i.e.~the leading term, is numerically dominant over non-perturbative contributions up to next-to-next-to leading order, both those suppressed by quark masses and those that are not.
\end{abstract}


\end{frontmatter}

\section{Introduction}

The experimentally measured value of the muon anomalous magnetic moment $a_{\mu} = (g-2)_{\mu}/2$ is~\cite{Bennett:2006fi,Tanabashi:2018oca}
\begin{equation}
a^{\mathrm{exp}}_{\mu}= 116\, 592\,  091 (63) \times 10^{-11} \, .
\end{equation}
The Standard Model (SM) prediction on the other hand is~\cite{Aoyama:2020ynm}
\begin{equation}
 a_{\mu}^{\mathrm{SM}} = 116\,  591\,  810  (43)  \times 10 ^{-11}\, ,
\end{equation}
i.e.~there is a $3.7\sigma$ discrepancy between theory and experiment. As a consequence, the muon $g-2$ is an excellent low energy observable for the hunt for new physics. With further improvement on precision it will be possible to deduce the nature of this discrepancy.

\begin{figure}[t!]\centering
	\includegraphics[width=0.25\textwidth]{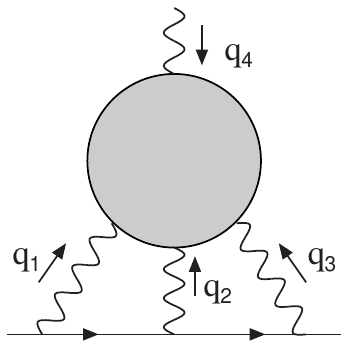}
	\caption{\label{fig:hlbl} A diagrammatic representation of the HLbL contribution to the $g-2$. The grey blob contains hadrons and the photon momenta have been labelled $q_{1,2,3,4}$, where the external photon is static, i.e.~$q_{4}\rightarrow 0$.}
\end{figure}

The SM prediction receives contributions from several sectors, namely quantum electrodynamics (QED), electroweak (EW) physics as well as the hadronic sector~\cite{Aoyama:2020ynm}. The bulk of the value comes from QED which is known very precisely, and the second most precise piece is the EW one. The hadronic sector is commonly divided into two pieces, the hadronic vacuum polarisation (HVP) and the hadronic light-by-light (HLbL). In numbers one has~\cite{Aoyama:2020ynm}
\begin{eqnarray}
&& a_{\mu}^{\mathrm{QED}} = 116 \, 584 \, 718.931(104)\times 10^{-11} \, ,
\\
&& a_{\mu}^{\mathrm{EW}} = 153.6(1.0)\times 10^{-11} \, ,
\\
&& a_{\mu}^{\mathrm{HVP}} = 6845(50)\times 10^{-11} \, ,
\\
&& a_{\mu}^{\mathrm{HLbL}} = 92(18) \times 10^{-11}\, .
\end{eqnarray}
The sum of these yields $a_{\mu}^{\mathrm{SM}} $. It is here clear that the hadronic contributions dominate the uncertainty and therefore require further consideration. Note that all higher order corrections to the HVP and HLbL here have been included in  $a_{\mu}^{\mathrm{HVP}}$ and $a_{\mu}^{\mathrm{HLbL}}$ even though we in the following will be interested in the leading order HLbL contribution, which is usually denoted $a_{\mu}^{\mathrm{HLbL},\,  \mathrm{LO}}$ and diagrammatically represented as in Fig.~\ref{fig:hlbl}. 

The HLbL is the time-ordered correlator of four electromagnetic (EM) currents, and for the $g-2$ requires loop integration over three virtual photons with momenta $q_{1}$, $q_{2}$ and $q_{3}$. The fourth, external, photon has momentum $q_{4}$ which for the $g-2$ is in the static limit $q_{4}\rightarrow 0$. The situation is depicted in Fig.~\ref{fig:hlbl}. In the following we will refer to the Euclidean photon virtualities $Q_{1,2,3}^2= -q_{1,2,3}^2$. The HLbL process can be calculated either in a data-driven approach using dispersion theory and models, or using lattice QCD, see Ref.~\cite{Aoyama:2020ynm} and references therein. The data-driven approach requires the knowledge of the HLbL tensor in different kinematic regions, i.e.~for different combinations of $Q_{i}^2$. The purely short-distance (SD) region is given by $Q_{1}^2\sim Q_{2}^{2}\sim Q_{3}^2 \gg \Lambda _{\mathrm{QCD}}^2$. We derive SD constraints (SDCs) for this particular kinematics by means of an operator product expansion (OPE) in the presence of an external static field corresponding to the external photon in Fig.~\ref{fig:hlbl}~\cite{Bijnens:2019ghy,Bijnens:2020xnl}. Such SDCs provide information on omitted higher order contributions in the dispersive approach and furthermore constrain model calculations. 

\section{Generalities about the HLbL tensor}

The HLbL tensor is a time-ordered correlation function of four EM currents. The currents are of the form $J^{\mu}(x) = \bar{q}\, Q_{q}\gamma ^{\mu}q$ where $q=(u,d,s)$ and $Q_{q}=\textrm{diag}(e_q) =\textrm{diag}(2/3,-1/3,-1/3)$ is the light quark charge matrix. With this one has the tensor as
\begin{eqnarray}\label{eq:hlbltensor}
\Pi^{\mu_{1}\mu_{2}\mu_{3}\mu_{4} } (q_{1},q_{2},q_{3}) = 
-i\int \frac{d^{4}q_{4}}{(2\pi)^{4}}\left(\prod_{i=1}^{4}\int d^{4}x_{i}\, e^{-i q_{i} x_{i}}\right) 
\hspace{-100pt}
\nonumber \\
\times \langle 0 | T\left(\prod_{j=1}^{4}J^{\mu_{j}}(x_{j})\right)|0\rangle \, .
\end{eqnarray}
The Ward identities for $q_{1,2,3,4}$, with $q_{4} = -q_{1}-q_{2}-q_{3}$, can be compactly written as
\begin{equation}
 q_{i,\, \mu_{i}} \, \Pi^{\mu_{1}\mu_{2}\mu_{3}\mu_{4}}(q_{1},q_{2},q_{3})=0  \, ,
\end{equation}
from which it follows that we may write~\cite{Aldins:1970id}
\begin{equation}\label{eq:widerres}
\hspace{-15pt}
\Pi^{\mu_{1}\mu_{2}\mu_{3}\mu_{4}}(q_{1},q_{2},q_{3})=-q_{4,\, \nu_{4}}\frac{\partial \Pi^{\mu_{1}\mu_{2}\mu_{3}\nu_{4}}}{\partial q_{4,\, \mu_{4}}}(q_{1},q_{2},q_{3}) \, .
\end{equation}
As a consequence, we may therefore calculate the HLbL by considering the derivative in~(\ref{eq:widerres}). Lorentz decomposing the HLbL tensor into a basis of 54 scalar functions $\Pi _{i}$ as in Refs.~\cite{Colangelo:2015ama,Colangelo:2017fiz} one has the HLbL contribution to $a_{\mu}$ as
\begin{eqnarray}\label{eq:amuhlblint}
\hspace{-5pt}
    a_{\mu}^{\mathrm{HLbL}} = \frac{2\alpha ^{3}}{3\pi ^{2}} 
     \int _{0}^{\infty} dQ_{1}\int_{0}^{\infty} dQ_{2} \int _{-1}^{1}d\tau \, \sqrt{1-\tau ^{2}}\,
     \hspace{-15pt}
    \nonumber \\
     \times \,  Q_{1}^{3}Q_{2}^{3} \sum _{i=1}^{12} T_{i}(Q_{1},Q_{2},\tau)\, \overline{\Pi}_{i}(Q_{1},Q_{2},\tau)\, .
\end{eqnarray}
Here, $T_i$ are known functions and the $\overline{\Pi}_{i}$ are functions of six linear combinations of the $\Pi _{i}$. The six linear combinations in question are commonly denoted as $\hat{\Pi}_{1,4,7,17,39,54}$, and together determine $a_{\mu}^{\mathrm{HLbL}}$.

As we are interested in the SD domain, we have to consider values of $Q_{i}$ greater than some $Q_{\mathrm{min}}$, which means that the integrals in~(\ref{eq:amuhlblint}) have a lower cut-off. Where to choose this cut-off is a priori not clear and is one of the sources of uncertainty in the HLbL prediction~\cite{Aoyama:2020ynm}. The determination of $Q_{\mathrm{min}}$ lies beyond our scope for the moment and we therefore let it be a variable quantity.

\subsection{Constraints from the short-distance domain}
As was explicitly shown in Ref.~\cite{Bijnens:2019ghy}, an OPE of the HLbL tensor in~(\ref{eq:hlbltensor}) does not lead to SDCs for the $g-2$ kinematics. The reason is that the systematic expansion breaks down due to the appearance of quark propagators of the soft momentum $q_{4}$. However, it is possible to consider the static external field as a background in which one then can construct a systematic OPE of a correlation function of three EM currents~\cite{Bijnens:2019ghy,Bijnens:2020xnl}. The quantity to study then is
\begin{eqnarray}
\label{eq:3pointem}
\Pi ^{\mu_{1} \mu_{2} \mu_{3} }(q_{1},q_{2}) = 
-\frac{1}{e}\int\frac{d^4 q_{3}}{(2\pi)^4}  \left(\prod_{i=1}^{3}\int d^{4}x_{i}\, e^{-i q_{i} x_{i}}\right) 
\hspace{-18pt}
\nonumber \\
 \times  \langle 0 | T\left(\prod_{j=1}^{3}J^{\mu_{j}}(x_{j})\right) | \gamma(q_4) \rangle \, .
\end{eqnarray}
Note that the static photon appears in the external state. The connection between $\Pi ^{\mu _{1} \mu _{2} \mu _{3}}$ and the HLbL tensor is obtained by factoring out the external field according to~\cite{Bijnens:2019ghy}
\begin{equation}
\hspace{-20pt}
\Pi^{\mu_{1}\mu_{2}\mu_{3}}(q_{1},q_{2})= \langle 0 |e_{q} F_{\nu_{4}\mu_{4}}|\gamma(q_{4})\rangle  
\times 
 \frac{i}{2}  \lim_{q_{4}\rightarrow 0}  \frac{\partial \Pi^{\mu_{1} \mu_{2} \mu_{3} \mu_{4}}}{\partial q_{4}^{\nu_{4}}}\, .
\end{equation}
An OPE here will therefore allow to obtain $\lim_{q_{4}\rightarrow 0}  \frac{\partial \Pi^{\mu_{1} \mu_{2} \mu_{3} \mu_{4}}}{\partial q_{4}^{\nu_{4}}}$ which is needed to find the $\hat{\Pi}_{i}$ and from these also $a_{\mu}^{\mathrm{HLbL}}$. 

\section{An OPE in an external EM field}
We want to construct an OPE for the tensor $\Pi ^{\mu _{1} \mu _{2} \mu _{3}}$, i.e.~where the external soft photon represents the background field. Such OPEs in the presence of an external field have been considered before, e.g.~for nucleon magnetic moments in Refs.~\cite{Balitsky:1983xk,Ioffe:1983ju} and also for the EW contributions to the muon magnetic moment in Ref.~\cite{Czarnecki:2002nt}. 

An OPE is a systematic expansion of a product of operators, and when applied to the EM currents in  $\Pi ^{\mu _{1} \mu _{2} \mu _{3}}$ it gives rise to a sum of contributions depending on non-perturbative matrix elements, corresponding to long-distance effects, multiplied by perturbative SD coefficients. In practice this sum is obtained by expanding the time-ordered product of operators in Dyson series and using Wick's theorem. The non-perturbative matrix elements come from not fully contracted terms in the Wick expansion. The resulting sum has a systematic ordering with long- and short-distance effects separated. Note that for each order one must analyse which operators can contribute for a given OPE. The main difference between a vacuum OPE and ours with a background field is that all operators which have the same quantum numbers as the external field $F_{\mu \nu}$ can contribute. 

We consider the OPE up to dimension $\alpha _{s}$ and $(\Lambda _{\mathrm{QCD}}/Q_{\mathrm{min}})^6$. There are eight types of operators which can contribute up to this order, namely
\begin{eqnarray}
&&
\hspace{-15pt}
 S_{1,\,\mu\nu}
=  e\,   e_{q}  F_{\mu\nu} 
\, ,\
\\ 
&& 
\hspace{-15pt}
S_{2,\,\mu\nu}
=  \bar{q}\sigma_{\mu\nu}q   \, ,
\\ 
&&
\hspace{-15pt}
 S_{3,\,\mu\nu}
=  i \,  \,\bar{q} G_{\mu\nu}q  \, ,
\\ 
&& 
\hspace{-15pt}
S_{4,\,\mu\nu}
=  i \, \bar{q} \bar{G}_{\mu\nu}\gamma_{5} q \, ,
\\
&& 
\hspace{-15pt}
S_{5,\,\mu\nu}
= \bar{q} q\; e\,   e_{q}F_{\mu\nu} \, ,
\\ 
&&
\hspace{-15pt}
 S_{6,\,\mu\nu}
=  \frac{\alpha_{s}}{\pi}\, G_ {a}^{\alpha\beta}G^{a}_{\alpha\beta}\; e\,   e_{q}F_{\mu\nu}  \, ,
\\
&&
\hspace{-15pt}
 S_{7,\,\mu\nu}
=  \bar{q}(G_{\mu\lambda}D_{\nu}+D_{\nu}G_{\mu\lambda})\gamma^{\lambda}q-(\mu\leftrightarrow\nu) \, ,
\\
&& 
\hspace{-15pt}
S_{\{8\},\,\mu\nu}
=  \alpha_{s}\, (\bar{q}\, \Gamma \,q \; \bar{q}\Gamma q)_{\mu\nu} \, .
\end{eqnarray}
We have here defined $G_{\mu\nu}= i g_S\lambda^a G^a_{\mu\nu}$ as well as its dual $\bar{G}^{\mu\nu} = \frac{i}{2}\epsilon^{\mu\nu\lambda\rho} G_{\lambda\rho}$, and $\Gamma$ is a matrix in spinor, colour and flavour space. For further conventions see Ref.~\cite{Bijnens:2020xnl}. Operators $ S_{\{8\},\,\mu\nu}$ involve flavour mixing, as they are obtained by expanding one order in $\alpha _{s}$ and non-contracted four-quark operators are obtained in the Wick expansion. In the chiral limit one can obtain a basis of twelve possible four-quark operators, as shown in Ref.~\cite{Bijnens:2020xnl}. 

\begin{figure}[t!]\centering
	\includegraphics[width=0.5\textwidth]{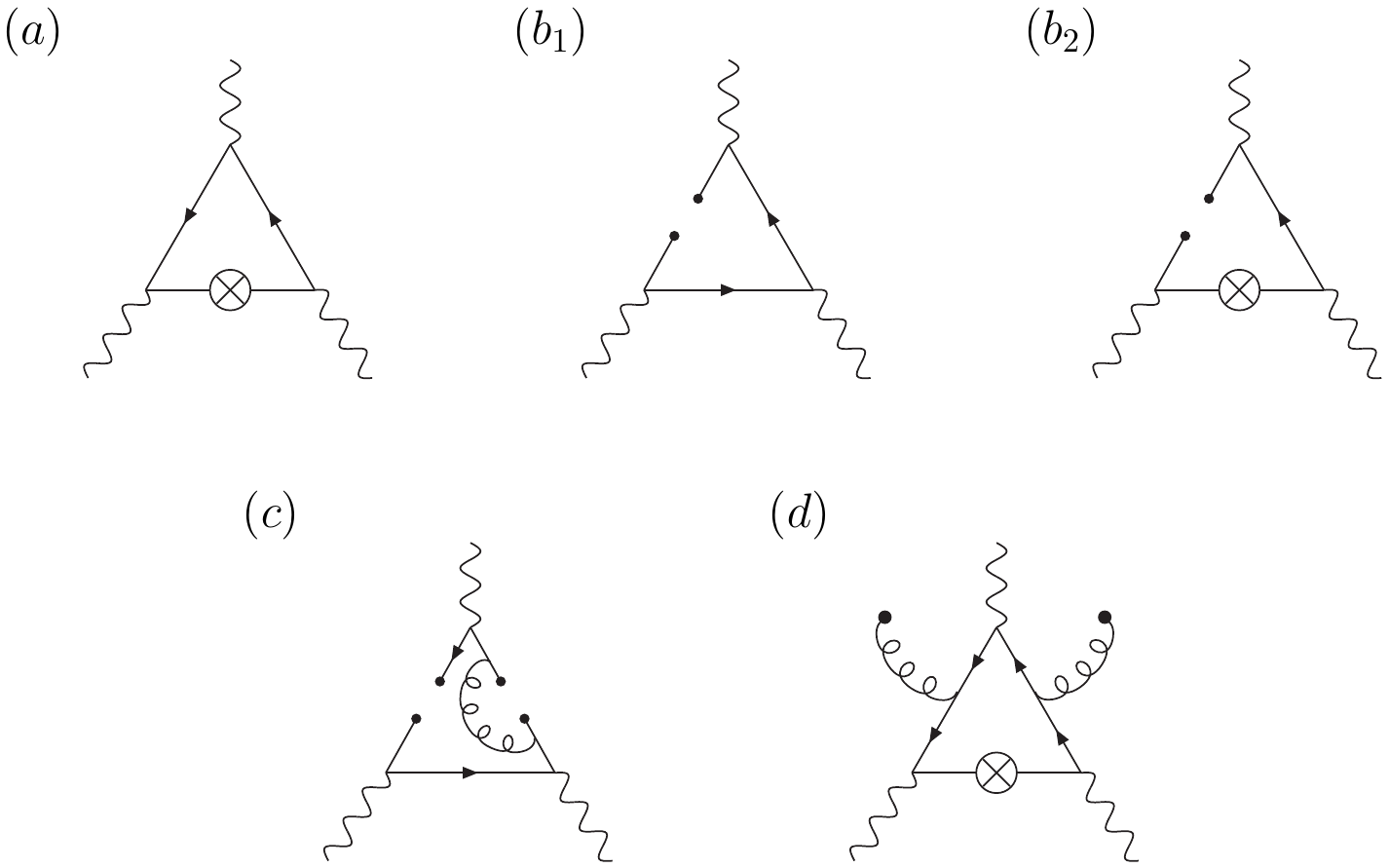}
	\caption{\label{fig:opediagrams} A diagrammatic representation of the  terms in the OPE up to dimension 6 and order $\alpha _{s}$. The external field can appear either on hard quark lines as an insertion vertex (marked by a cross) or in a non-perturbative matrix element (in diagrams where the cross has been omitted). The terms are ($a$) the quark loop, ($b_{1}$) the di-quark magnetic susceptibility $\langle \bar{q} \sigma _ {\mu \nu} q$, ($b_{2}$) the quark condensate $\langle \bar{q}q\rangle $, ($c$) four-quark condensates $\langle \bar{q} \Gamma _{1} q \, \bar{q}\Gamma _{2} q\rangle $ for colour, flavour and Dirac matrices $\Gamma _{1,2}$, and ($d$) the gluon condensate $\langle \alpha _{s}GG\rangle $.  }
\end{figure}

The diagrams appearing in our OPE are shown in Fig.~\ref{fig:opediagrams}. This first of all shows that the quark loop is the first term in a systematic OPE. This has for a long time been a common assumption lacking an explicit derivation, and it has further been assumed that it is a sufficiently good representation of the SD HLbL behaviour~\cite{Bijnens:2019ghy,Bijnens:2020xnl}. Note that we so far have not calculated the leading perturbative $\alpha _{s}$ correction to the quark loop, but this is the subject of an upcoming publication.

\subsection{Renormalisation}

Simply following the above procedure for the OPE is not sufficient. The result for $\Pi^{\mu_{1}\mu_{2}\mu_{3}}$ can be written
\begin{equation}\label{eq:operes}
\hspace{-15pt}
\Pi^{\mu_{1}\mu_{2}\mu_{3}}(q_{1},q_{2}) = \vec{C}^{\, T,\, \mu_{1}\mu_{2}\mu_{3}\mu_{4}\nu_{4}}(q_{1},q_{2})\,\vec{X}_{S}\,\langle e_{q}F_{\mu_{4}\nu_{4}}\rangle \, ,
\end{equation}
where the vector $\vec{X}_{S}$ has components $X_{S}^{i}$ defined through $\langle S_{i,\, \mu\nu}\rangle = e \, e_q X_{S}^{i} \langle F_{\mu\nu} \rangle$. In other words, the components $X_{S}^{i}$  are related to the magnetic susceptibilities of the respective operators. The vector $\vec{C}^{\, T,\, \mu_{1}\mu_{2}\mu_{3}\mu_{4}}$ contains the momentum dependence. At this stage, however, long and short distances are not yet completely separated. This can be understood since from non-contracted soft quark lines there are contributions from the Dyson series giving rise to a divergent series. In addition, in~(\ref{eq:operes}) one obtains corrections $\log Q_{i}^2/m_{q}^2$ that lead to ill-defined series. We solve these issues by dressing and renormalising the operators, in the $\overline{MS}$ scheme, as explained further in Ref.~\cite{Bijnens:2020xnl}. After including operator mixing and defining dressed and renormalised operators $\vec{Q}_{\overline{MS}, \, \mu\nu}(\mu)$ at scale $\mu$, the result in~(\ref{eq:operes}) takes the form
\begin{equation}\label{eq:operesren}
\hspace{-20pt}
\Pi^{\mu_{1}\mu_{2}\mu_{3}}(q_{1},q_{2}) = \frac{1}{e}\vec{C}^{\, T,\, \mu_{1}\mu_{2}\mu_{3}\mu_{4}\nu_{4}}_{\overline{MS}}(q_{1},q_{2})\langle \vec{Q}_{\overline{MS}, \mu_{4}\nu_{4}}(\mu)\rangle
 \, ,
\end{equation}
where
\begin{equation}
\label{eq:magnsusc}
\langle\vec{Q}_{\overline{MS}, \, \mu\nu}(\mu)\rangle =
e\, \vec{X} \, \langle e_{q}F_{\mu\nu}\rangle \, .
\end{equation}
The magnetic susceptibilities corresponding to the renormalised operators are here contained in the vector. Here long and short distances have been completely separated and no divergent mass logarithms appear. To conclude, compared to the massless quark loop we have calculated contributions of order $g_{s}\,\frac{\Lambda_{\mathrm{QCD}}^4}{Q^4}$, $\frac{m_q^2}{Q^2}$, $g_s^2 \,\frac{\Lambda_{\mathrm{QCD}}^4}{Q^4}$, $m_q\frac{\Lambda_{\mathrm{QCD}}}{Q^2}$, $m_q\frac{\Lambda_{\mathrm{QCD}}^3}{Q^4}$ and $m_q^3\frac{\Lambda_{\mathrm{QCD}}}{Q^4}$.

\subsection{Numerical estimates for the matrix elements}
Having performed the renormalised OPE to the desired order, the only remaining step to calculate $a_{\mu}^{\mathrm{HLbL}}$ is to determine the magnetic susceptibilites in~(\ref{eq:magnsusc}). First of all, $X_{5}$ and $X_{6}$ are respectively given by the quark and gluon condensates. The quark condensate is well-studied and its numerical value can be found in many places, e.g.~Ref.~\cite{Aoki:2019cca}. The gluon condensate was estimated in Ref.~\cite{Shifman:1978bx} to be $X_{6}\sim 0.02\, \mathrm{GeV}^{4}$. For $X_{7}$ we simply make a guess inspired from the operator mixing, namely $|X_{7}|\sim\frac{1}{6}\Big\langle \frac{\alpha_{s}}{\pi}GG\Big\rangle $. For the four-quark operators only two combinations appear, which can be estimated using large-$N_{c}$ arguments. The values coincide and are $\overline{X}_{8,1}^{N_c\rightarrow \infty}
= \overline{X}_{8,2}^{N_c\rightarrow \infty}
=-2\frac{\pi\alpha_{s}}{9}X_{2}\langle \bar{q}{q}\rangle$. The matrix element $X_{2}$ is the so-called di-quark magnetic susceptibility of the vacuum and has been calculated on the lattice, see Ref.~\cite{Bali:2020bcn} for a recent value. The only remaining matrix elements are $X_{3}$ and $X_{4}$ which to our knowledge were hitherto unknown. In order to estimate them we connect the matrix elements to vacuum QCD two-point functions and employ large-$N_{c}$ arguments. The results are $X_{3}=-\frac{m_{0}^{2}}{6M_{\rho}^2}\langle \bar{q} q\rangle $ and $
X_{4}=-\frac{m_{0}^{2}}{6M_{\rho}^2}\langle \bar{q} q\rangle $, where $m_{0}^2$ is a parameter estimated in Ref.~\cite{Belyaev:1982sa}. Using the same approach for $X_{2}$ yields $X_{2}=\frac{2}{M_{\rho}^2}\langle\bar{q}q\rangle $ which is in excellent agreement with lattice determinations. For further details on the numerical values of the matrix elements and analytic expressions from~(\ref{eq:operesren}), see Ref.~\cite{Bijnens:2020xnl}.

\section{Numerical results for $a_{\mu}^{\mathrm{HLbL}}$}

\begin{figure*}[t!]
\begin{minipage}{.48\textwidth}
  \centering
  \includegraphics[width=1.\textwidth]{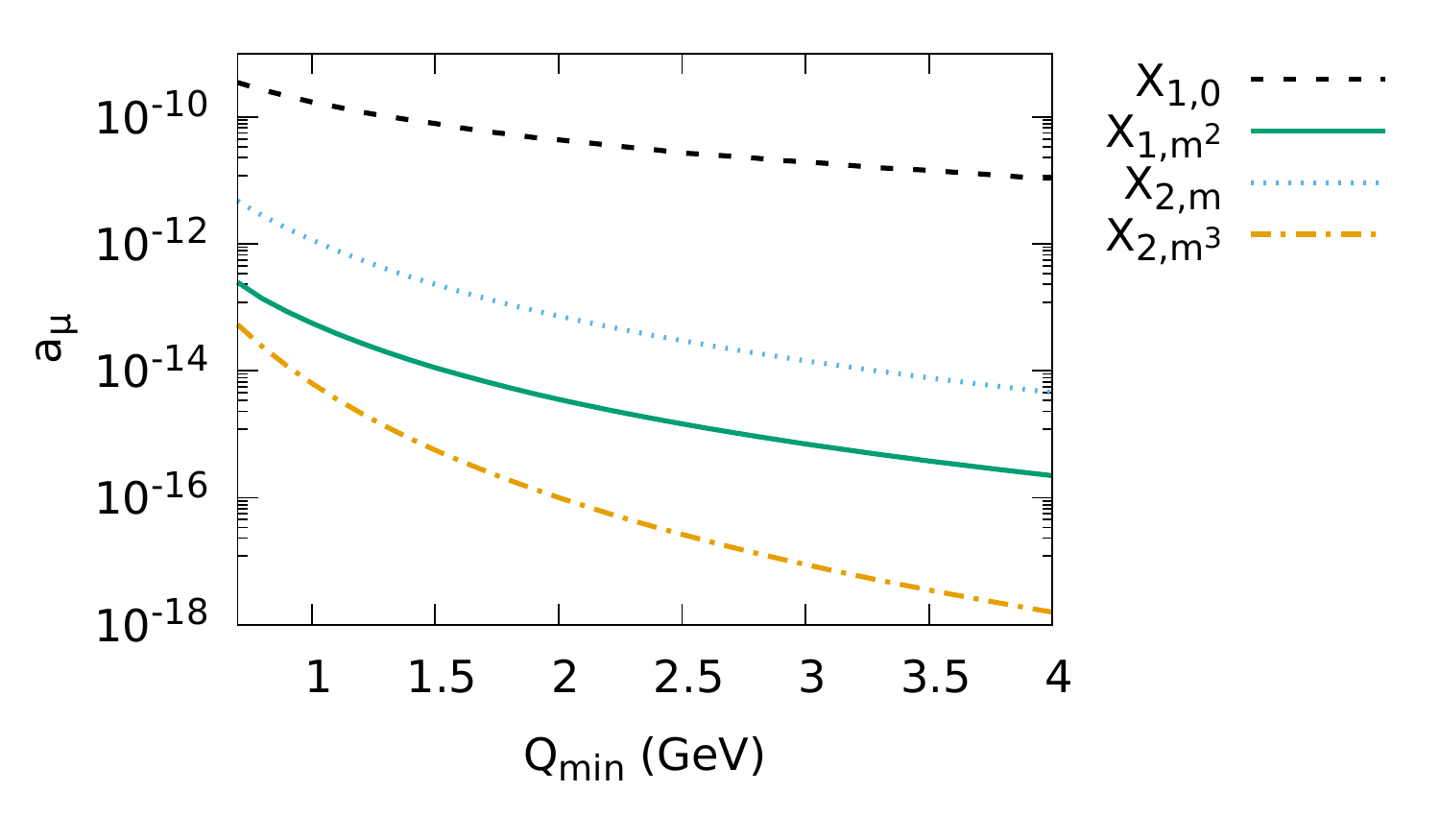}
	\caption{Contributions to $a_{\mu}^{\mathrm{HLbL}}$from $X_{1,0}$, $X_{1,m^2}$, $X_{2,m}$ and $X_{2,m^3}$. }
	\label{fig:plot1}
\end{minipage}
\begin{minipage}{.48\textwidth}
  \centering
 \includegraphics[width=1.\textwidth]{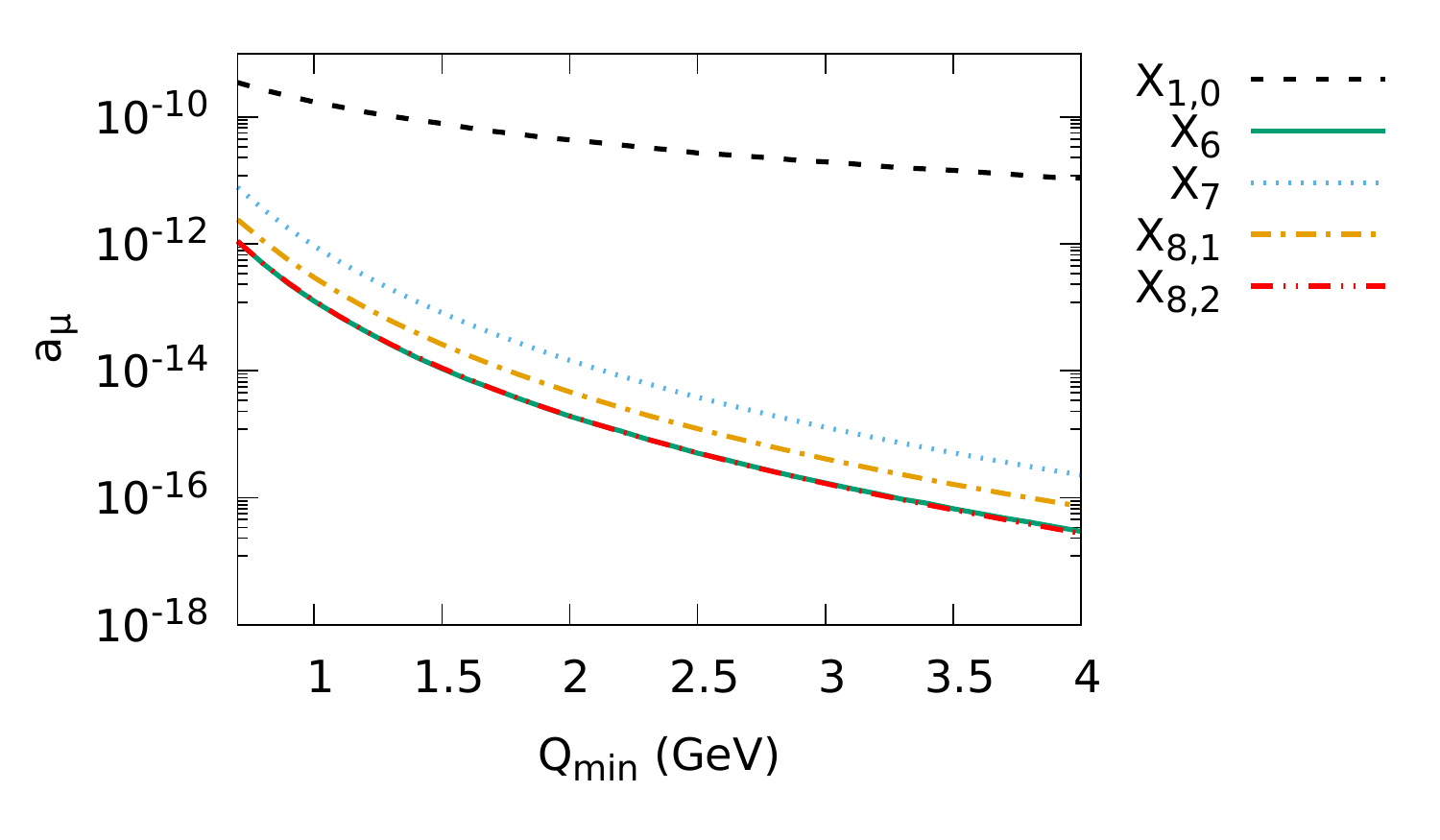}
\caption{Contributions to $a_{\mu}^{\mathrm{HLbL}}$from $X_{1,0}$, $X_{6}$, $X_{7}$, $X_{8,1}$ and $X_{8,2}$. }  
  \label{fig:plot2} 
\end{minipage}
\end{figure*}
\begin{figure*}[t!]
\centering
 \centering \includegraphics[width=0.48\textwidth]{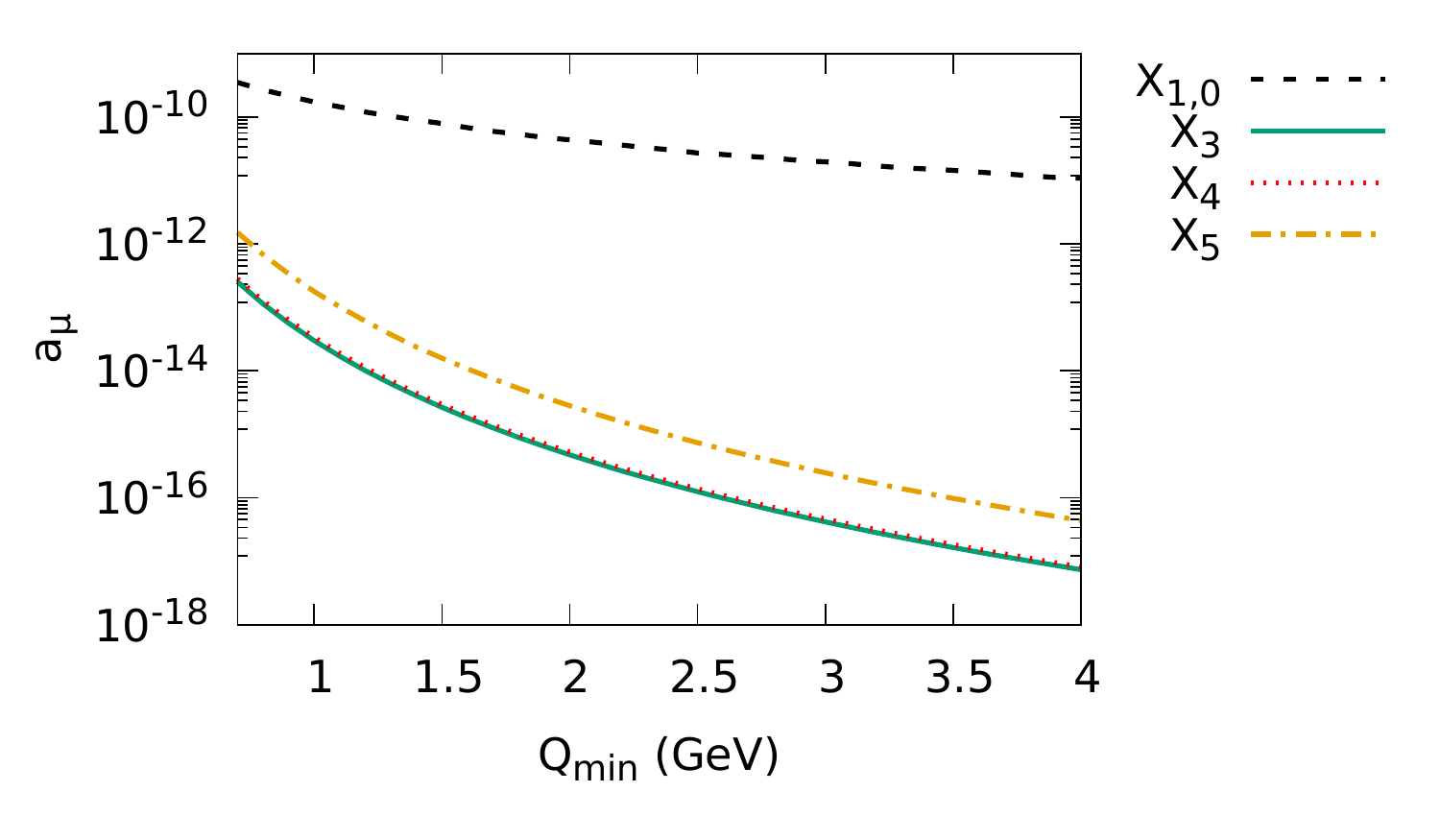}
	\caption{Contributions to $a_{\mu}^{\mathrm{HLbL}}$from $X_{1,0}$, $X_{3}$, $X_{4}$ and $X_{5}$. }
	\label{fig:plot3}
\end{figure*}

In Figs.~\ref{fig:plot1}--\ref{fig:plot3} we plot $a_{\mu}^{\mathrm{HLbL}}$ in~(\ref{eq:amuhlblint}) as a function of $Q_{\mathrm{min}}$. We have here used $\alpha _{s}=0.33$, $m_{u} = m_{d} = 5$ MeV, $m_{s} = 100$ MeV and the renormalisation scale $\mu = Q_{\mathrm{min}}$. The massless quark loop given by $X_{1,0}$ clearly dominates with respect to the other contributions to the OPE. This is true for both the contributions suppressed by quark masses such as e.g.~$X_{2,m}$ as well as those that are not such as the four-quark pieces $X_{8,1}$ and $X_{8,2}$. As a consequence, the massless quark loop seems to be a good representation of the SD HLbL behaviour from relatively low energies. However, before such a statement can be made, one has to study also the $\alpha _{s}$ correction to the massless quark loop, i.e.~by including two extra quark-gluon-antiquark QCD vertices in the Dyson series. Our current preliminary evaluation shows that also this is small compared to the massless quark loop. 

\section{Conclusions}
We have developed a systematic OPE in the presence of an external EM field to derive SDCs for the HLbL tensor. These constraints are needed to reduce the error on the SM prediction of the muon anomalous magnetic moment. We have considered the non-perturbative contributions up to dimension $6$ and order $\alpha _{s}$. We have shown that the massless quark loop is the first term in this OPE, thus putting a long-standing assumption on firm theoretical ground. We have also shown that this leading term dominates over the non-perturbative corrections, both those suppressed by quark masses and those that are not. The only piece remaining to finally deduce how good a representation the massless quark loop is of the SD HLbL tensor is the order $\alpha _{s}$ corrected massless quark loop. Our preliminary findings show that also these corrections are small, which will shortly be presented in an upcoming publication. 

\section*{Acknowledgements}
N.~H.--T. and L.~L.~are funded by the Albert Einstein Center for Fundamental Physics at Universit\"{a}t Bern and the Swiss National Science Foundation respectively. J.~B.~and A.~R.--S.~are supported by the Swedish Research Council grants contract numbers 2016-05996 and 2019-03779. A.~R.--S.~is partially supported by the Agence Nationale de la Recherche (ANR) under grant ANR-19-
CE31-0012 (project MORA).

\bibliographystyle{elsarticle-num}
\bibliography{refsqcd20}

\end{document}